# Graphone (one-side hydrogenated graphene) formation on different substrates

C. F. Woellner [a], P. A. S. Autreto [b] and D. S. Galvão [a]

a. Applied Physics Department, State University of Campinas, Campinas-SP, 13083-970, Brazil. E-mail: galvao@ifi.unicamp.br
b. Centro de Ciências Naturais e Humanas, Universidade Federal do ABC, Santo André, 09210-580 SP, Brazil. E-mail: pedro.autreto@ufabc.edu.br

In this work we present a fully atomistic reactive (ReaxFF force field) molecular dynamics study of the structural and dynamical aspects of the one-side hydrogenation of graphene membranes, leading to the formation of the so-called graphone structure. We have considered different substrates: graphene, few-layers graphene, graphite and platinum at different temperatures. Our results showed that the hydrogenation rates are very dependent on the substrate and thermal effects. Our results also showed that, similarly to graphane, large hydrogenated domains are unlikely to be formed. These hydrogenation processes occur through the formation of uncorrelated cluster domains.

## Introduction

The chemistry of carbon is very rich and this richness is due, mainly, to the fact that there are three different orbital hybridizations (sp, $sp^2$ and $sp^3$). This characteristic allows a plethora of distinct allotropes, some of them discovered and/or experimentally realized in the last few decades, such as nanotubes, fullerenes[1] and one of the most important subject in materials science today, graphene[2].

Graphene is a two-dimensional array of hexagonal units of $sp^2$ bonded carbon atoms, which has been theoretically investigated since late 1940s as a model to describe some properties of graphite-based materials[3]. Graphene (single layer graphite) was obtained in 2004 by Novoselov and Geim[2], using the "scotch tape" method. They were able to obtain single layer graphene from highly oriented pyrolytic graphite (HOPG). Since then, its unusual and extraordinary electronic and mechanical properties continue to attract the attention of the scientific community.

Considering applications in the field of nanoelectronics[4–8], this material is also of great importance. This is due in part to the fact that conduction electrons in this material can be described like two-dimensional fermions with zero rest mass (zero gap) and travelling at high speeds, typically of the order of $10^6$ m/s. These and other properties make graphene an interesting material for future electronics. Even though this material presents several remarkable properties, there are some difficulties to be overcome before a real graphene-based nanoelectronics becomes a reality. These difficulties are mainly related to its electronic zero bandgap value that, despite of being intrinsically interesting for the conical shape and for providing relativistic behavior for conduction electrons, hinders its direct use for some nanodevices, like digital transistors and diodes.

Many strategies have been tried to overcome the gapless character of graphene[9–11]. These strategies include exploring physical and chemical methods, as for example; quantum dots[12], strain[13], nanoribbons[14–16], chemical functionalization as hydrogenation[17–19], oxidation[20–22] and more recently, fluorination[23,24]. Actually, the chemical functionalization approach appears to be the most promising one, since it can be used either to open the electronic gap, as well as to directly modify the interactions of graphene with its environments. This can be exploited in many and different technological applications, even for graphene immersed into liquids, for example. Functionalization can also be important for other varied graphene aspects, such as: drug delivery, hydrogen storage[17,25], defect manipulation[25,26], magnetic effects[25–27], reduction of graphene oxide to produce large scale and inexpensive graphene[28], protection of nanoribbons edges[26], functionalization to crack graphene in order to produce pieces in specific shape[29], among others[30].

The fully hydrogenated graphene form, the so-called graphane, was proposed by Sofo and co-authors in 2007[17] and consists of a single-layer structure of carbon atoms with C-H bonds corresponding to a weakly coupled diamond-like layers, but with the carbon atoms in $sp^3$ hybridizations, instead of $sp^2$ ones. Different groups have already reported the graphane-like experimental realization[19,31]. More recently, Elias et al.[19] demonstrated the existence of graphane-like structures formed by exposing both (up and down) graphene membrane sides to hydrogen plasma.

Especially interesting is the graphene semi-hydrogenated (just one hydrogenated side) form, the so-called graphone. Predicted by Zhou and co-workers[32], graphone is a also a semiconductor, as graphane, but in addition it may exhibit magnetic properties[32–34], as for example, ferromagnetic behavior at room temperature[32]. This can occur because when half of the carbon atoms are hydrogenated, the π-bonding network is disrupted, leading to localized and unpaired electrons in the non-hydrogenated carbon atoms[35]. The magnetic ordering (ferromagnetic, antiferromagnetic) as well as the bandgap values can even be controlled by the hydrogenation patterns. Those features could make graphone a promising material for future spintronics applications[36].

Despite graphone sheets can be synthesized in a couple of different ways, one-side hydrogenation of supported/suspended graphene is the most common one. In this case, the influence of the substrate has been proven to be crucial[36–38] in the resulting hydrogenation patterns, which affects directly the magnetic properties[34,39,40]. However, a detailed understanding of these processes is lacking and to address these issues is one of the objectives of the present work.

In this work, we have investigated through fully atomistic reactive molecular dynamics (MD) simulations the dynamics and structural patterns of the hydrogenation processes of one-side graphene membranes supported on three different substrates: graphene (thus forming a bilayer graphene structure), graphite and platinum, on the configurations shown in Figure 1. For comparison purposes, we have also considered the cases of one-side hydrogenated suspended graphene membranes.

## Computational Methods

In order to study the effects of the substrates on the processes of the one-side graphene hydrogenation, we have carried out simulations using fully atomistic reactive molecular dynamics methods (ReaxFF)[41], as available in the open source code Large-scale atomic/ Molecular Massively Parallel Simulator (LAMMPS)[42]. ReaxFF is a force field developed by van Duin, Goddard III and co-workers. It allows an accurate description of chemical processes, such as the formation and breaking of chemical bonds, without taking explicitly into account the electronic part, as in fully quantum methods. Due to this it has a relatively low computational cost, which allow handling systems containing many thousand atoms. As other non-reactive force fields, such as MM3[43], the total energy is divided into several energy terms, as bond stretching, bond angle bending, van der Waals and Coulomb interactions, among others. The ReaxFF parameterization is obtained using Density Functional Theory (DFT) calculations and/or experimental data and its accuracy, in comparison to the experimental data, is around 2.9 kcal/mol for unconjugated and conjugated systems. This method has been successfully applied in many investigations of nanomaterials at the atomic scale, including chemical functionalizations[9,24,44]. For more details see reference 41.

All MD simulations were performed under a NVT ensemble and using Noosé-Hover thermostat[45] for controlling the temperature. Typical time simulation was 1 ns, with time steps of 0.1 fs. Three different temperatures were considered: 450, 550 and 650K to determine how the hydrogenation dynamics depends on thermal effects.

As mentioned above in addition to suspended graphene (Figure 1a), we have considered graphene on three different substrates: graphene (Figure 1b), forming a bilayer system, graphite (Figure 1c), and platinum (Figure 1d). Suspended graphene was simulated using a rim. For graphene on graphene (bilayer system), we kept all atoms free to move and use again a rim as a structural support. For all configurations, hydrogenation was allowed to occur only on the center of the structures, in order to avoid spurious edges effects (see Figure 1).

The MD simulations were carried out considering the different membranes and/or configurations immersed into an atmosphere of atomic hydrogen atoms. For speeding up the simulations the hydrogen-hydrogen atom interactions were turned off. The typical dimension simulation box was of 150 Å x 150 Å x 60 Å, containing a membrane of dimensions of 150 Å x 150 Å with the hydrogen accessible area being of 100 Å x 100 Å. The hydrogen atmosphere was composed of 1000 atoms.

In order to gain further insights on the preferential hydrogenation sites and its dependence on the substrates, we have also calculated the potential energy maps. These maps were calculated considering a potential energy variation of a probe hydrogen atom located 1.5 Å above the graphene basal plane for each membrane configuration (isolated and deposited on substrates). Negative values indicate energetically favorable regions for hydrogen atom reactions.

## Results and discussion

The interactions between substrate and graphene are predominantly non-covalent, mainly through van der Waals forces. Despite weak, the interactions can produce changes in the geometry of structure. They can also change the local graphene chemical reactivity and, consequently, the ratio and pattern of the hydrogenation processes. These effects can be better visualized from the potential energy maps, which can also provide relevant information about preferential sites for hydrogenation.

These potential energy maps are presented in Figure 2, for suspended graphene and also for graphene membranes deposited on the three different substrates. Although the substrate interactions are not very strong, they are still stronger enough to affect the maps.

For the suspended graphene the typical pattern consists of C-C bonds as attractive for hydrogenation and the centers of benzene rings are well-defined non-attractive areas. This equivalence symmetry for the isolated graphene is broken by the presence of the substrates. From Figure 2 we can see that maps are different from the corresponding suspended graphene case. Thus, we can expect that the hydrogenation processes will be also affected. This is indeed corroborated by the MD simulations.

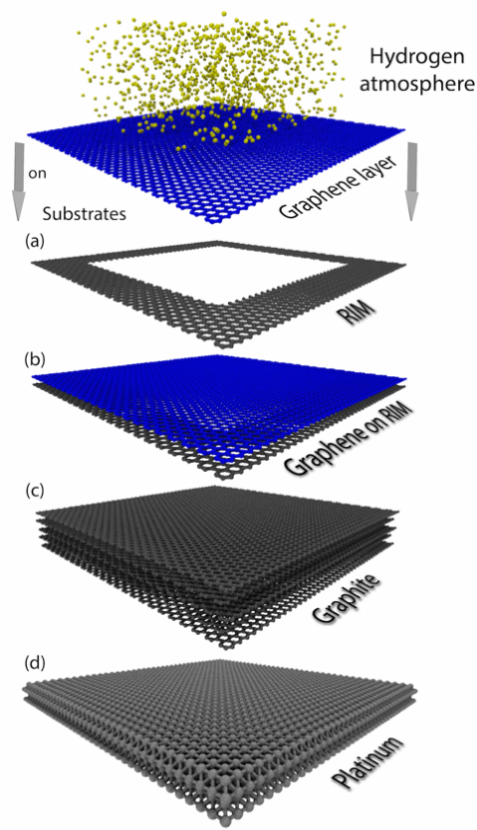

**Figure 1:** Configurations investigated in the present work: (a) suspended graphene; (b) graphene on graphene (bilayer system); (c) graphite and; (d) platinum.

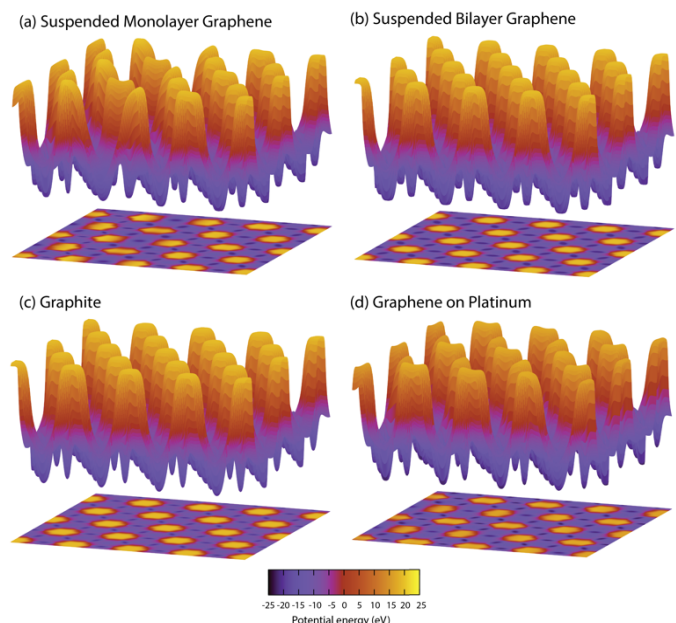

**Figure 2.** Potential energy maps (ReaxFF results) for a potential experienced by hydrogen atom probe placed at a distance of 1.5 Å above the: (a) suspended graphene plane; (b) graphene on graphene (bilayer system); (c) graphite and; (d) platinum.

In Figure 3 we present the MD results for the number of bonded hydrogen atoms per available carbon (H/C), as a function of time for three different temperatures; 450, 550 and 650K, respectively. As expected, the results show that hydrogenation is highly dependent on temperature values. For 450K, less than 5% of carbons are functionalized whereas this ratio reaches more than 40% for 650K (only 150K higher). For all cases considered here, the order of most functionalized structures were temperature independent. Graphite and bilayer graphene (BLG) compose a group of poorly functionalized, followed by single layer graphene (SLG) and graphene on platinum. The last one is the most functionalized structure, consistent with the fact that the use of metallic substrates improves the hydrogenation effectiveness.

One important result is that in the graphone formation it is not easily distinguishable the two common regimes observed for graphene hydrogenation[9], i. e.; fast adsorption of hydrogen atoms followed by an almost linear hydrogenation regime. This can be explained by the fact that in the case of having both membrane sides available for hydrogenation, when a hydrogen bond is formed the induced structural deformations makes the next neighbor carbon atom very reactive from the other membrane side. The presence of the substrate precludes this, thus making the total hydrogenation processes slower.

The hydrogen bond formation significantly alters the potential energy around its carbon neighbors (see Figure 4a, for the case of platinum substrate) making this region more favorable to further hydrogenation. This results in a hydrogenation processes occurring via non-correlated island domains, which was also observed for graphane[9]. In Figure 4 we present four representative MD snapshots of the graphene hydrogenation on platinum showing this hydrogen cluster formation. As we can see from this Figure, the hydrogen incorporation occurs with the nucleation and growth of independent (uncorrelated) domains. This is a substrate independent phenomenon.

The influence of substrate in hydrogenation process is due to dynamics effects driven by van der Waals interactions. In Figure 5, we highlight the geometry of graphene after the first hydrogen atom is attached. As we can see, the buckling for graphene on graphene and on graphite substrates is larger than in suspended and platinum cases. The level of buckling follows the same hydrogenation ratio effectiveness. This is a direct consequence of the substrate interactions. Graphene and graphite substrates do not favor hydrogen attachments close to carbon atoms already hydrogen bonded. This results in carbon atoms with hybridizations close to sp$^2$. On the other hand suspended and platinum substrates, exhibiting a higher level of hydrogenation, result in carbon atoms with more sp$^3$-like hybridizations.

Another relevant result is that in graphone formation the hydrogenation occurs through growing domains (islands of hydrogenated carbons), which has been also reported in the case of graphane formation[9]. As mentioned above this is a consequence of breaking the symmetry of C-C conjugated bonds by the first attached hydrogen atom which favors the hydrogenation of their neighbors. The quite different dynamics for the different cases can be better evaluated in the movies of the supplementary materials.

Figure 5 also shows the final stage after hydrogenation simulation at 650K. Even for this temperature and with a significant number of bonded atoms, the structural integrity of the membrane is yet not compromised, in contrast to the observed in fluorographene formation[24]. This is in good agreement with previous experimental and theoretical published works[32]. Our saturation hydrogenation level is also in good agreement with the available experimental data that show that it is very difficult to achieve high hydrogenation levels. This contrasts with 100% hydrogenation assumed in many idealized models reported in the literature[19,32,36,46] and stress the importance of considering large and disordered systems in order to obtain realistic model of graphene functionalization.

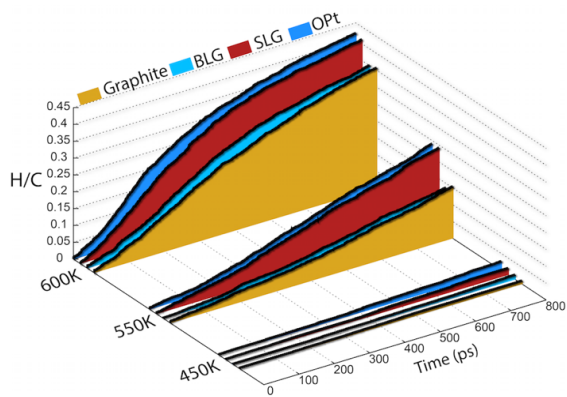

**Figure 3.** Relative (H/C) hydrogenation rate as function of time for different temperatures

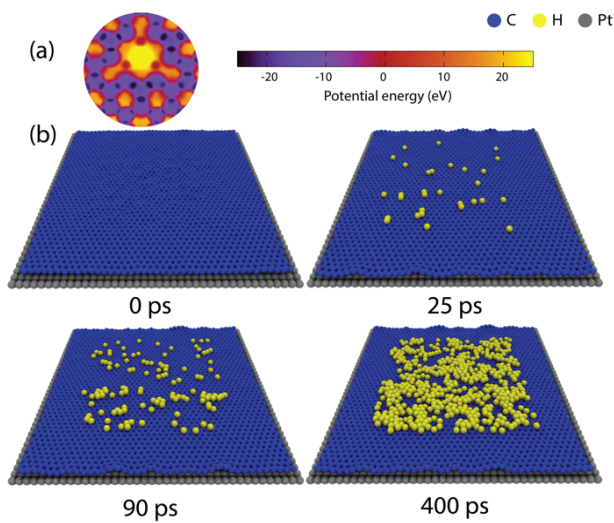

**Figure 4:** Hydrogen cluster formation: (a) Potential energy map experienced by a hydrogen atom probe at 1.5 Å around an already formed C-H bond on a graphene membrane deposited on a platinum substrate; (b) Four representative MD snapshots showing the hydrogen cluster formation and growth.

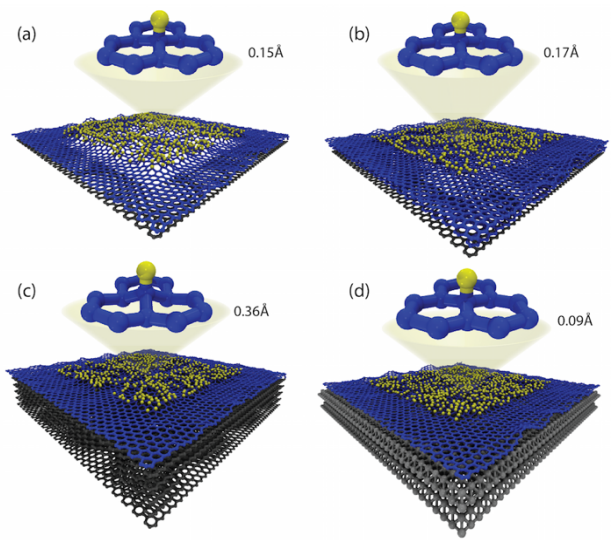

**Figure 5:** Final stage of one-side hydrogenated graphene (graphone) on the different substrates. In the inset is show the first hydrogen bond formation.

## Summary and Conclusions

We have carried out a fully atomistic reactive molecular dynamics study on the dynamics of hydrogenation of one-side graphene membranes (graphone) on different substrates and at different temperatures. Our results have showed that graphene hydrogenation rate is very sensitive to the substrate type. Platinum and suspended graphene have showed the largest hydrogenation coverage. As expected, the hydrogenation coverage is proportional to temperatures. The existence of two hydrogenation regimes (fast hydrogen absorption followed by an almost linear increase) reported to graphane does not occur for graphone. Our results also show that during the early stages of the hydrogenation a significant number of randomly distributed and uncorrelated domains (islands) are formed. These results are substrate and temperature independent. These findings suggest that similarly to graphane, large perfect graphone-like domains are unlikely to be formed. Thus, the use of graphone in spintronics applications, as has been speculated[36], requiring magnetic ordering will be very difficult, unless in a very diluted regime.

## Acknowledgements

CFW thanks São Paulo Research Foundation (FAPESP) Grant No. 2014/24547-1 for financial support. The authors acknowledge computational and financial support from the Center for Computational Engineering and Sciences at Unicamp through the FAPESP/CEPID Grant No. 2013/08293-7.

## Notes and references